\documentclass[twocolumn,showpacs,preprintnumbers,amsmath,amssymb,prb]{revtex4}
\usepackage{graphicx}
\usepackage{dcolumn}
\usepackage{bm}
\usepackage{comment}

\begin{document}

\preprint{APS/123-QED}

\title{ Spin relaxation in inhomogeneous quantum dot arrays studied by electron spin resonance}

\author{A.~F.~Zinovieva}%
\email{aigul@isp.nsc.ru}
\author{N.~P.~Stepina}
\author{A.~I.~Nikiforov}
\author{A.~V.~Nenashev}
\author{A.~V.~Dvurechenskii}
\affiliation{%
Institute of Semiconductor Physics, SB RAS, 630090 Novosibirsk,
Russia
}%
\author{L.~V.~Kulik}
\affiliation{%
Institute of Chemical Kinetics and Combustion, SB RAS, 630090
Novosibirsk, Russia
}%

\author{N.~A.~Sobolev}
\author{M.~C.~Carmo}
\affiliation{%
Departamento de Fisica e I3N, Universidade de Aveiro, Aveiro,
Portugal
}%

\date{\today}

\begin{abstract}
{Electron states  in a inhomogeneous Ge/Si quantum dot array with
groups of closely spaced quantum dots  were studied by conventional
continuous wave ($cw$) ESR and spin-echo methods. We find that the
existence of quantum dot groups  allows to increase the spin
relaxation time in the system. Created structures allow us to change
an effective localization radius of electrons by external magnetic
field. With the localization radius close to the size of a quantum
dot group, we obtain fourfold increasing spin relaxation time $T_1$,
as compared to conventional homogeneous quantum dot arrays. This
effect is attributed to averaging of local magnetic fields related
to nuclear spins $^{29}$Si and stabilization of $S_z$-polarization
during electron back-and-forth motion within a quantum dot group.

}
\end{abstract}

\pacs{73.21.La, 03.67.Lx, 72.25.Rb}

\maketitle

\section*{Introduction}

Electron spins in quantum dots (QDs) can be considered as promising
candidates for realization of quantum computation ideas and
spintronics devices. \cite{a1,a2} The main parameter indicating the
applicability of a system for quantum computation is the spin
coherence time. An extremely long spin lifetime is observed in
zero-dimensional structures due to a strong confinement in all three
dimensions. \cite{a3} An especially great potential for a long
coherence time is expected in the Ge/Si system with quantum dots. In
this system electrons are localized in strained Si regions, where
spin-orbit (SO) coupling is very weak. However, recent
investigations of spin decoherence by spin echo method in the Ge/Si
QD system \cite{a4} demonstrated that the spin relaxation times are
unexpectedly short ($\sim10\mu s$). It was suggested that the reason
of such intensive spin relaxation consists in the appearance of
effective magnetic fields during electron tunneling between quantum
dots. These magnetic fields (Rashba fields)\cite{a5} originate from
spin-orbit interaction and arise due to the absence of mirror
symmetry of the localizing potential for an electron in the vicinity
of Ge QD. Spin relaxation occurs through stochastic spin precession
in effective magnetic fields during random tunneling between QDs
(analog of Dyakonov-Perel mechanism for delocalized
carriers\cite{a6}). Obviously, suppression of the tunneling in the
array of well-separated quantum dots allows to eliminate the
existence of in-plane fluctuating magnetic fields. In this case the
hyperfine interaction with $^{29}$Si nuclear spins comes into force
and determines the spin relaxation time. If the tunneling is
suppressed not by spatial separation of QDs, but by Coulomb
repulsion \cite{a7}, the anisotropic exchange interaction can also
control the spin relaxation process.

The efficiency of each mechanism depends in different ways on the
localization degree of electrons. Changing the tunnel coupling
between quantum dots (by changing their density) and
correspondingly, the localization degree of electrons, it is
possible to alter the relative contribution of different mechanisms.
With the increase of electron localization radius the contribution
of hyperfine interaction becomes smaller due to averaging-out of
different orientations of nuclear spins. A related increase of the
relaxation time occurs until the moment when the wave function
overlapping provides the hopping between neighboring localization
centers. In these conditions Dyakonov-Perel mechanism begins to
control the spin relaxation process.\cite{a8} Longest spin
relaxation time is expected right before the point where
Dyakonov-Perel mechanism comes into force. Similar effect was
detected for n-type GaAs impurity system, where threefold increase
of spin relaxation time was obtained  in the vicinity of
metal-to-insulator transition.\cite{a9}

In self-assembled tunnel-coupled QD structures it is hard to get a
gradual change of the localization radius by changing the QD array
density. The stochastic nucleation of QDs during growth in
Stranskii-Krastanov mode \cite{a10}  leads to the formation of the
regions with a high local density of QDs. In such regions a strong
tunnel coupling between dots results in the intensive spin
relaxation through Dyakonov-Perel mechanism. However, under certain
conditions, the existence of groups of closely located QDs can
provide not a decrease, but an increase of the spin relaxation time.
First, QD groups should be well separated from each other. In this
case, the effective magnetic fields can be averaged due to electron
back-and-forth motion within each QD group. Second, QDs inside the
group should have strong tunnel coupling providing the effective
localization radius comparable with  QD group size. As a result, the
averaging of local magnetic fields related to nuclear spins will
take place.

The present work is devoted to the electron spin resonance (ESR)
study of inhomogeneous QD arrays, where the averaging of Rashba and
hyperfine fields inside QD groups  is expected to provide a long
spin relaxation time. We succeed in creating the experimental
structure containing well separated groups of QDs with a large
electron localization radius. The coupling between QDs and,
consequently, the electron localization radius in the structures
under study turned out to be dependent on the external magnetic
field orientation. A fourfold increase of the spin relaxation time
as compared to the previous data for dense homogeneous QD arrays
\cite{a4} has been detected at a special orientation of magnetic
field, where the electron localization radius was close to the QD
group size.

\section {Samples and Experiment}

The samples were grown by molecular-beam epitaxy on n-Si$(001)$
substrates with the resistivity of 1000 $\Omega\,$cm. To increase
the response from the sample, we have grown 6 layers of the Ge
nanoclusters separated by 30 nm Si-layers. Each QD layer was formed
by deposition of 7 ML Ge at the temperature $T=550^{\circ}$C. On the
top of the structure, a $0.3\,\mu$m epitaxial n-Si layer (Sb
concentration $\geq10^{17} $cm$^{-3}$) was grown, the same layer was
formed below QD layers. The scanning tunneling microscopy (STM) of
the structure with a single QD layer uncovered by Si shows the
bimodal distribution of QDs ($hut$- and $dome$-clusters) (Fig.~1).
The density of $dome$-clusters is $\sim10^{10}$~cm$^{-2}$, the
typical base width is $l=50$~nm, the height is $h=10$~nm. The
$hut$-clusters are distributed between $dome$-clusters with density
$\sim10^{11}$~cm$^{-2}$, their typical base width is $l=15$~nm, the
height is $h=1.5$~nm.

The obtained sizes of $dome$-clusters have to provide a strong
localization of electrons at the apex of such type QDs with a small
localization radius. To increase the localization radius, we used
the temperature 500$^{\circ}$C for overgrowth of QDs allowing to
transform $dome$-clusters to disk-like clusters without intensive
Ge-Si intermixing inside QDs.  The cross-section images obtained by
TEM ( transmission electron microscopy) show that the height of
disk-like dots does not exceed 3 nm in the experimental structure.
These dots, as well as original $domes$, are characterized  by the
absence of mirror symmetry due to a difference between the smeared
top and sharper bottom of QDs. So, after such overgrowth the
localization radius is expected to be comparable with the lateral
size of QD.

STM data show a non-homogenous in-plane distribution of
$dome$-clusters and the existence of groups of 2-3 closely spaced
nanoclusters on the average (Fig.~1). A sufficient tunneling
coupling between them allows the electron wave function to spread
over the group of QDs and promotes a further increase of the
electron localization radius.

$Hut$-clusters in these structures can not be the centers of
localization because the binding energy of electron on such type
quantum dots is very small ($\sim10$meV).\cite{a11} Recently, to
provide the localization of electrons on $hut$-clusters, the stacked
structures with four layers of Ge quantum dots were grown
\cite{a12}. The distances between QD layers were 3~nm, 5~nm, and
that resulted in the effective deepening of potential well near
$hut$-clusters due to accumulation of strain from different QD
layers. In the structure under study the distance between QD layers
is 30 nm, then the strain accumulation does not occur.

\begin{figure}
\includegraphics[width=2.8in]{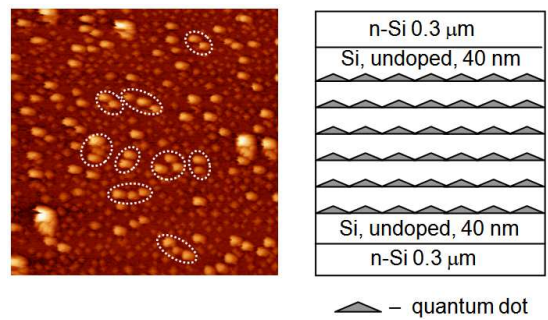}
\caption{\label{f1} Right panel: STM of the uncovered sample with
two-shaped QDs ($hut$-clusters and $dome$-clusters), 1
$\mu$m$\times$1 $\mu$m image. Left panel: a schematic structure of
investigated sample. The examples of quantum dot groups are
indicated by dashed line loops.}
\end{figure}

ESR measurements were performed with a Bruker Elexsys~580 X-band EPR
spectrometer using a dielectric cavity Bruker ER-4118 X-MD-5. The
samples were glued on a quartz holder, then the entire cavity and
the sample were maintained at a low temperature with a helium flow
cryostat (Oxford CF935). The needless EPR signal from dangling bonds
($g=2.0055$) was avoided using the passivation of structures with
atomic hydrogen before measurements. To increase the number of
registrable spins, the sandwiched sample was prepared. Samples were
thinned by acid etching up to $150-250\,\mu$m. After thinning,
samples were glued together; finally, the object composed of 4-5
wafers was investigated.

The spin echo measurements were carried out at temperature 4.5~K in
resonance magnetic field $H=3470$~G (can be slightly varied $\pm5$~G
depending on resonance conditions) with direction corresponding to
the narrowest ESR line width, $\theta=30^{\circ}$, where $\theta$ is
the angle between magnetic field and growth direction of the
structure [001]. A two-pulse Hahn echo experiment
($\pi/2\!-\!\tau\!-\!\pi\!-\!\tau\!-$~echo) was used to measure
$T_2$ (a detailed explanation can be found in
Ref.~\onlinecite{a13}). In order to observe a longitudinal spin
relaxation (corresponding time $T_1$), a different pulse sequence is
applied ($\pi\!-\!\tau\!-\!\pi/2\!-\!T\!-\!\pi\!-\!T\!-$~echo). The
first $\pi$-pulse rotates the magnetization opposite to its thermal
equilibrium orientation, where the interaction with the environment
causes the spins to relax back to the initial orientation parallel
to $\mathbf{H}$. After time $\tau$, a $\pi/2$-pulse followed by
another $\pi$-pulse is used to observe a Hahn echo. In the first and
second type of experiments, the durations of $\pi/2$ and $\pi$
pulses were 60~ns and 120~ns, respectively; the interpulse time in
the second experiment was kept $T=200$~ns.

\section {Results}

The ESR spectra measured at different directions of magnetic field
are shown in Fig.~2, where $\theta=0^{\circ}$ corresponds to the
magnetic field applied parallel to the growth direction $Z$. At
$\theta=0^{\circ}$ the ESR line is most symmetrical and its shape is
close to Gaussian. The line asymmetry becomes more pronounced with
the increase of angle $\theta$ and the line shape tends to
Lorentzian already at $\theta=10^{\circ}$. The line shape analysis
performed at $\theta=30^{\circ}$ is shown  in Fig.~3.  A careful
examination of ESR line shape shows that the ESR line represents the
sum of an absorption line (dotted) and a dispersion line (dashed).
The rotation of the sample in the magnetic field results in a change
of the resonance line width and the resonance field. The orientation
dependence of the ESR line width  for the structure under study is
demonstrated in Fig.~4 . When the external magnetic field deviates
from the growth direction up to $\theta\approx30^{\circ}$, the ESR
line width sharply decreases from $\Delta H$=1.9~Oe to $\Delta
H$=1.4~Oe. A further tilt of the magnetic field leads to the line
broadening with maximum $\Delta H$=2.4~Oe at $\theta=60^{\circ}$.
For the in-plane magnetic field, the ESR line width is narrowed
again down to $\Delta H$=1.8~Oe. Such nonmonotomic behavior is
unusual for electrons in 2D system, and has not been observed up to
now.

The angular dependence of g-factor is shown in Fig.~5. At small
angles (up to $30^{\circ}$) the g-factor slightly changes nearly
$g=1.9994(5)$. Between $\theta=30^{\circ}$ and $\theta=40^{\circ}$
the g-factor value jumps to $g=1.9992$ and remains nearly constant
up to $\theta=90^{\circ}$.

 \begin{figure}
\includegraphics[width=3.3in]{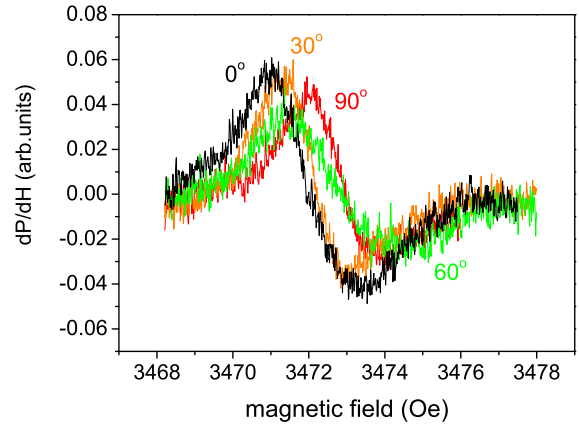}
\caption{\label{f2}  ESR spectra at different orientations of
magnetic field. For $\theta=0^\circ$ the magnetic field is parallel
to the growth direction of the structure [001], $\theta=90^\circ$
corresponds to magnetic field applied along crystallographic
direction [110].}
\end{figure}

\begin{figure}
\includegraphics[width=2.8in]{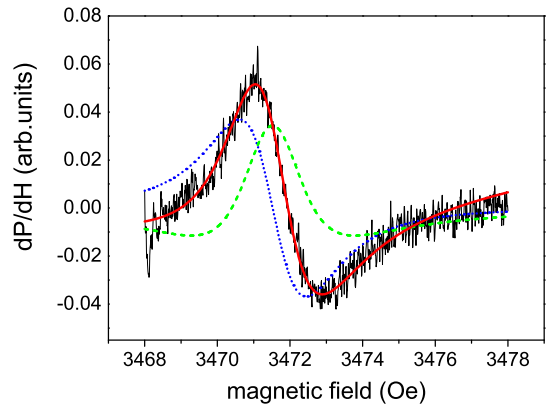}
\caption{\label{f6}  Analysis of the ESR line shape for
$\theta=30^{\circ}$. The solid line represents the sum of an
absorption line (dotted), and a dispersion line (dashed).}
\end{figure}

\begin{figure}
\includegraphics[width=3.2in]{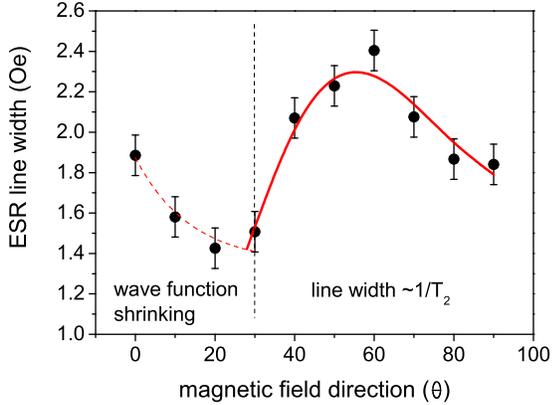}
\caption{\label{f64} The angular dependence of ESR line width and
theoretical approximation (Eq.6)  of this dependence (solid line)
for the structure under study. For $\theta=0^\circ$ the magnetic
field is parallel to the growth direction of the structure. }
\end{figure}

\begin{figure}
\includegraphics[width=3.2in]{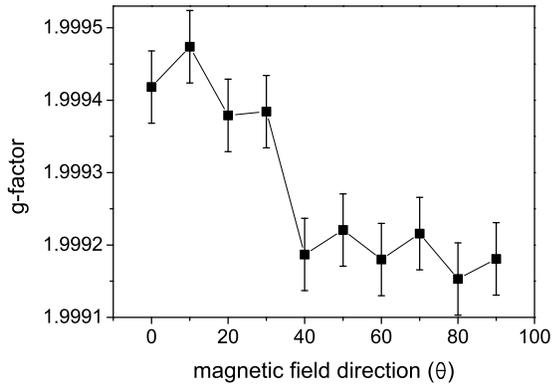}
\caption{\label{f25}  The angular dependence of electron g-factor for
the structure under study. For $\theta=0^\circ$ the magnetic field
is parallel to the growth direction of the structure.}
\end{figure}

The data of spin echo measurements performing at
$\theta=30^{\circ}$, when the most narrow  ESR line width is
observed, are shown in Fig.~6,~7.
\begin{figure}
\includegraphics[width=3.5in]{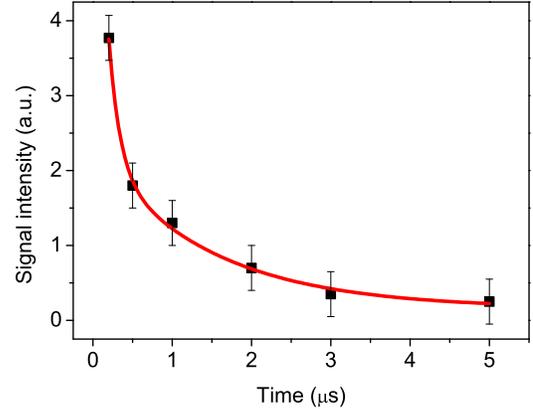}
\caption{\label{f4}  Results of two-pulse spin echo experiments
performed at $\theta=30^\circ$  (points) and approximation by
superposition of two exponential functions, Eq.1 (solid line).
Corresponding microwave pulse sequence is $\pi/2-\tau
-\pi-\tau-echo$.}
\end{figure}
According to the results of a two-pulse Hahn echo experiment, the
spin echo behavior can be described by superposition of two
exponentially decaying functions:
\begin{eqnarray}
M(t)=M^{(1)}_{x,y}\,\exp(-2\tau/T_2^{(1)})+M^{(2)}_{x,y}\,\exp(-2\tau/T_2^{(2)}),
\end{eqnarray}
where $M(0)=M^{(1)}_{x,y}+M^{(2)}_{x,y}$ is the lateral (in QD
plane) magnetization after $\pi/2$-pulse. The decay parameters  give
two times of spin dephasing: $T_2^{(1)}\approx0.26 ~\mu$s and
$T_2^{(2)}\approx1.5~\mu$s.

The analysis of the inversion signal recovery measured in
three-pulse echo experiments shows  non-exponential behavior
(Fig.~7). The experimental curve can be described by the
superposition of two functions:
\begin{eqnarray}
 M(t)=M_{0z}-M^{(1)}_z \exp(-\tau/T_1^{(1)})-M^{(2)}_z \exp(-\tau/T_1^{(2)}),
\end{eqnarray}
where $M_{0z}$ is the equilibrium magnetization,
$M_{0z}=M^{(1)}_{0z}+M^{(2)}_{0z}$,
$M^{(1,2)}_z=M^{(1,2)}_{0z}-M^{(1,2)}_z(0)$,
$M_z(0)=M^{1}_z(0)+M^{2}_z(0)$ is the magnetization just after
applying of an inverting $\pi$-pulse. In correspondence with this
equation at the beginning the samples magnetization recovers very
fast. After some time the part of spins is returned to an
equilibrium state and the recovery rate becomes much slower.  The
characteristic times obtained by fitting the experimental data are
$T_1^{(1)}\!\approx\!2\,\mu s$ and $T_1^{(2)}\!\approx\!35\,\mu$s.
All values of spin relaxation times were determined with error
$\pm20\%$.

\begin{figure}
\includegraphics[width=3.5in]{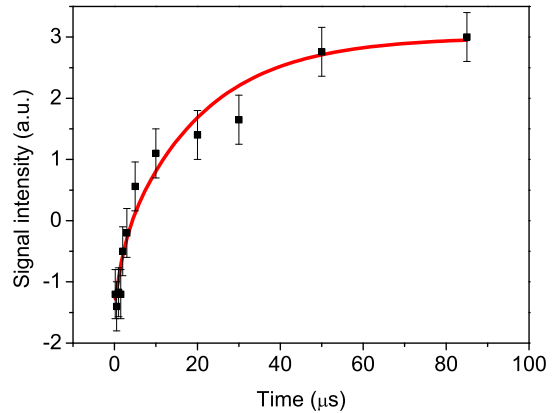}
\caption{\label{f47}  Amplitude of inversion-recovery signal versus
interpulse delay $\tau$ (symbols  are experimental data, solid line
is the approximation by Eq.2). Corresponding microwave pulse
sequence is $\pi-\tau -\pi/2-T-\pi-T-echo$. Experiments are
performed at $\theta=30^\circ$.}
\end{figure}

\section {Discussion}

To explain the experimental results obtained in the present work, we
propose the following model (see Fig.~8). Electrons are suggested to
localize mainly in the groups of closely spaced QDs containing 2-3
QDs on the average (see STM data). During overgrowth QDs lose their
apexes and transform to the disk-like shape QDs. As a result, the
electron localization radius can become comparable with QD lateral
size. Additional barriers for electrons limiting the electron motion
in $XY$ directions in the Si layer arise due to the regions with a
nonzero Ge content. These regions are located over the edges along
the perimeter of QDs and they are formed following the mechanism of
formation of SiGe rings described in Ref.\onlinecite{a14}.  Thus,
the electron localization radius can be taken about 50 nm for a
spatially isolated single QD. In the case of a group of closely
spaced QDs the separating SiGe barriers between dots inside the
group are absent because of energetically unfavorable positions of
Ge atoms between QDs due to a high strain.\cite{a15} SiGe barriers
remain only along the external border of QD groups. Then the
electron wave function can spread to the size of the QD group,
$l\sim100-150$ nm. Since the confinement of electrons is not too
strong, the tails of electron wave functions from different QD
groups can be overlapped providing the hopping between QD groups.
The external magnetic field applied along the growth direction can
sufficiently change the described picture. Magnetic length
$\lambda=\sqrt{c\hbar/eH} $,  in our experimental set up (H=3470
Oe), is about 45 nm, that is comparable with the electron
localization radius for a single QD. In these conditions, the
magnetic field effectively shrinks the tails of electron wave
functions, resulting in the enhancement of electron localization.
Thus, the perpendicular magnetic field suppresses the transitions
between QD groups and decreases the electron localization radius in
the QD group down to the size of an individual QD. Nevertheless, the
transitions between QDs inside the groups still persist due to a
small distance between QDs. With the deviation of the magnetic field
from the growth direction the probability of electron transitions
between dots increases and the conductivity in local areas (QD
groups) becomes higher. In experiment, this corresponds to the
appearance of a noticeable dispersion signal and the enhancement of
asymmetry of ESR line (ESR line becomes close to Dysonian line
\cite{a16}). A similar effect was observed for SiGe/Si/SiGe
structures with two-dimensional (2D) electron gas.\cite{a17}

At the same time the increase of the effective electron localization
radius  causes the averaging of the local magnetic fields induced by
nuclear spins and the smoothing of the QD parameter differences
within a QD group. As a result, the narrowing of ESR line at the
deviation of magnetic field from $\theta=0^{\circ}$ to
$\theta=30^{\circ}$ is observed. The minimum of ESR line width at
$\theta=30^{\circ}$ indicates that the electron localization radius
reaches the size of QD group and the full averaging inside each
group takes place. A further increase of the electron localization
radius leads to the enhancement of hopping between groups and a
decrease of spin lifetime through Dyakonov-Perel spin relaxation
mechanism.

The decreasing spin relaxation time affects the ESR line width. From
$\theta=30^{\circ}$ the broadening due to spin relaxation, $\Delta
H\cong1/T_2$, exceeds the nonhomogeneous broadening and further
orientation dependence of ESR line width is controlled by spin
relaxation time.

It should be noted that there is one more mechanism which can
provide the anisotropy of ESR line width, the relaxation assisted by
spin-phonon interaction. This mechanism can be effective due to the
lack of phonon bottleneck in our structures with large QD sizes
resulting in small confinement energies of electrons. In this case
the anisotropy of spin relaxation processes is defined by the shape
asymmetry of disk-like QDs. Their lateral size is one order larger
than their height, therefore only $k_x$- and $k_y$-phonon waves
effectively influence the spin. However, the experimentally observed
maximum of ESR line width at $\theta=60^{\circ}$ cannot be described
in the frames of the spin-phonon interaction model,\cite{a18} which
should provide a monotonic orientation dependence of ESR line width.

Dyakonov-Perel mechanism allows to describe non-monotonic behavior
of ESR line width on the assumption of $\tau_h$ depending on the
magnetic field. This dependence, determined in the frame of the
hopping model, \cite{a19} can be described as exponential:
\begin{eqnarray}
 \tau_h=\tau_0\exp(\alpha H_z^m),
\end{eqnarray}
where $H_z$ is the projection of magnetic field to the growth
direction; coefficient $m$ can be equal to $m=1/2$ or $m=2$ for case
of strong or weak magnetic fields. For intermediate fields
$\lambda\sim l$ this coefficient can take a value in the range of
$\{\frac{1}{2}\div2\}$ (Ref.~\onlinecite{a19}).

Spin relaxation time $T_2$ in the frames of Redfield theory
\cite{a20} is given by the following expression:
 \begin{eqnarray}
\frac{1}{T_2}=\gamma^2\delta H^{2}_{y}\sin^2\theta
\tau_h+\frac{1}{2T_1},
\end{eqnarray}
with
\begin{eqnarray}
\frac{1}{T_1}=\gamma^2(\delta H^{2}_{x}+\delta
H^{2}_{y}\cdot\cos^2\theta
)\frac{\tau_h}{1+\omega^{2}_{0}\tau^2_h},\nonumber\\
\end{eqnarray}
where correlation time of spin-orbit field fluctuations $\tau_c$ was
replaced by hopping time $\tau_h$; the  $\omega_0$ is the Larmor
frequency;  $\delta H_x$, $\delta H_y$ are the components of
effective magnetic field $\delta H$.

So, using expression (3) for $\tau_h$, we obtain the following
expression describing the orientational dependence of ESR line
width:
\begin{eqnarray}
\Delta{H}=B\exp(A \cos^m \theta)\left(\sin^2\theta+
\frac{1+\cos^2\theta}{1+C\exp(2A\cos^m\theta)}\right),
\end{eqnarray}
where $B=\gamma\delta H^{2}_{y}\tau_0$, $A=\alpha H^m$,
$C=\omega_0^2\tau_0^2$. The experimental data $\Delta H(\theta)$ in
the range of $\theta\in\{30^{\circ};90^{\circ}\}$ are  well
approximated by this expression (Fig.~4) with $m=3/2$, $A=1.52$,
$B=1.78$, $C=795.2$. Obtained coefficient $m$ corresponds to the
case of intermediate magnetic fields ($\lambda\sim l$) that argues
for accepted hopping model with $\tau_h$ depending on the magnetic
field.

\begin{figure}
\includegraphics[width=3.2in]{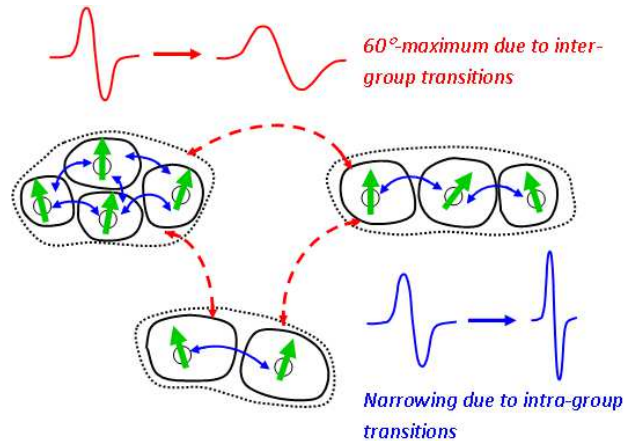}
\caption{\label{f28}  Illustration of a hopping model for the
structure under study. Hopping transitions inside the QD groups
provide the narrowing of ESR line with the deviation of magnetic
field from $\theta=0^{\circ}$ to  $\theta=30^{\circ}$. Hopping
transitions between QD groups provoke the spin relaxation through
the spin precession mechanism and give a special orientation
dependence of ESR line width in the range of
$\theta\in\{30^{\circ}-90^{\circ}\}$. }
\end{figure}

The magnitude of the effective magnetic field  $\delta H$, estimated
from the $B$ coefficient turns out to be $\approx15$~Oe. This value
is twice smaller than that determined in our previous work
\cite{a21} for $hut$ clusters with the aspect ratio $h/l$=0.1. It is
known \cite{a22} that in a QD system the effective magnetic field
depends on $h/l$, the more the aspect ratio the larger the $\delta
H$. For QDs under study the aspect ratio is about of $\approx0.05$,
therefore the effective magnetic field proved to be smaller.

The spin echo data are in a good agreement with the proposed model
of the existing  closed groups of quantum dots being the centers of
electron localization in the sample. Experimental spin polarization
behavior shows that the spin relaxation occurs in two stages, rapid
and slow ones. To understand the origin of this two-stage spin
dynamics we simulated the spin relaxation process in the ring-shaped
group of quantum dots. Model includes the strong tunneling coupling
between quantum dots in the circle. Hopping between any neighboring
QDs is permitted with an equal probability for back and forth
motion. Each tunneling transition is accompanied by spin rotation on
a small fixed angle $\alpha=0.1$. The direction of the rotation axis
is defined by product $[\mathbf{n}\times \mathbf{e_z}]$, where
$\mathbf{n}$ -tunneling direction, $\mathbf{e_z}$ -growth direction
of QD. The external magnetic field is applied along $\mathbf{e_z}$
and provides the Larmor precession between tunneling events. The
time intervals between tunneling events are distributed
exponentially with a mean value $\tau_h$.  The spin relaxation
caused by the interaction with phonons and nuclear spins was not
included into the consideration. The transport was simulated by
Monte-Carlo method for a different number of QDs in the circle.  The
results of simulation for the ring constructed of 10 quantum dots
are demonstrated in Fig.~9. The two stage spin dynamics is clearly
seen. It turned out that this effect depends on the relation between
hopping time $\tau_h$ and Larmor frequency $\omega_l$. The two-stage
dynamics is observed when $\omega_l \tau_h \ll 1$. For example, the
data in Fig.~9 were obtained at $\omega_l \tau_h = 0.1$. The first
stage of spin relaxation  is related to the processes of electron
spreading all over the group of QDs.  At this stage the loss of spin
polarization occurs due to the precession in the effective magnetic
field during tunneling between dots. The spin dynamics at the second
stage is defined by the phase breaking of Larmor precession during a
random walk along the QD ring. Generally speaking, this stage of
spin relaxation can be ruled by spin-phonon or hyperfine interaction
as well, if one includes them in the consideration. The absence of
two-stage dynamics in the case of $\omega_l \tau_h \geq 1$ can be
understood  by the simple consideration of spin behavior in the
frame of reference rotating with Larmor frequency. The randomness of
hopping between dots leads to averaging of effective magnetic field
$(\langle \delta H\rangle=0)$ and elimination of spin relaxation at
the first stage of electron extension over the group of QDs.
According to the simulation results, the first rapid stage is
characterized by a special relation between the longitudinal and
transverse spin relaxation times $T_1$ and $T_2$, usual for 2D
system with the absence of mirror symmetry $T_2\approx2T_1$.  Such
relation was obtained by spin echo measurements for 2D electron gas
structures \cite{a23}  and for dense homogeneous QD arrays,\cite{a4}
and follows from the in-plane arrangement of fluctuating magnetic
fields $\delta H$. Also we have verified the presence of two-stage
spin dynamics in QD clusters with other spatial arrangements, for
example, QD lines containing a few dots. The  described general
features are well preserved with changing only numerical values of
$T_1$ and $T_2$.

\begin{figure}
\includegraphics[width=3.2in]{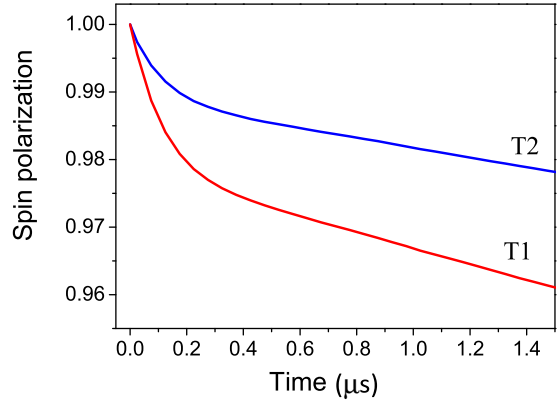}
\caption{\label{f29}  Results of spin relaxation simulation in a
closed ring-shaped group of QD. Number of QDs in the ring - $n=10$.
The hopping time is taken as $\tau_h=10^{-11}$s, Larmor frequency
was one order smaller $\omega=10^{10}$s$^{-1}$. Two stage dynamics
is observed. }
\end{figure}

In experiment on the $T_2$-measurements the first stage is
characterized by $T_2\approx0.26 ~\mu$s. Based on the simulation
results one can expect the same shortness for $T_1$. However, the
three-pulse method has limitations on measurements of such short
times. The difference between durations of the pulse sequences in
$T_1$-experiments and $T_2$-experiments is comparable with duration
of the first rapid stage of spin relaxation that makes difficult the
study of the beginning of $S_z$-relaxation.

The second stage of spin relaxation has the times $T_1=2\,\mu$s and
$T_2=1.5\,\mu$s. Here, the special relation $T_2\approx2T_1$ is not
fulfilled because of the presence of some additional spin relaxation
mechanisms, for example, Larmor precession phase breaking during a
random walk along the QD group or another one. The presence of long
living spin polarization with characteristic time
$T_1\approx35\,\mu$s is attributed to some stabilization of the
$S_z$-component taking place at the electron movement within a
closed QD group. According to the simulation results at a high
hopping frequency ($\omega_l \tau_h <0.01$) the spin polarization
after rapid stage is settled at some level depending on parameters
of QD groups. In this case, Larmor precession can be neglected, and
the sequence of small turnings in Rashba fields can be considered as
the effective precession around growth direction $Z$. In these
conditions, the $S_z$-component is stabilized. In contrast, the
transverse component of spin relaxes quickly, then in the experiment
we did not observe the rest transverse spin polarization with a long
relaxation time.

The orientational dependence of the g-factor  allows us to add some
details to the considered model. The value of g-factor g=1.9994,
within the experimental error, coincides with  typical g-factor
value for electron states  near the conduction band edge in
Si.\cite{a24} The fact that the g-factor remains near this value up
to $\theta\approx30^{\circ}$ confirms that electron is located in Si
regions until this orientation of magnetic field. In other words,
the electron increases its localization radius remaining in Si
regions. After point $\theta=30^{\circ}$, the electron localization
radius exceeds the size of QD groups and the g-factor drastically
changes to value $g=1.9992$.  Such behavior can be explained by the
penetration of electron wave function in SiGe regions surrounding QD
groups. The presence of Ge atoms can provide a decrease of the
g-factor value.\cite{a21}

It should be noted that the same value of electron g-factor was
obtained by us in another ESR experiment for the structure with
large SiGe nanodisks having diameter 100-150 nm. For this structure
we use substrate with specially created nucleation sites to obtain
more ordered array of quantum dots. These nucleation sites
originated due to strain modulation in the surface layer induced by
previously buried QDs. Large $dome$-clusters grown at  previous
stage at temperature $650^{\circ}$ have a good spatial ordering due
to a long range elastic interaction between QDs.\cite{a15} On this
strain-modulated surface we have grown  10 layers of QDs using the
same temperature regime as in the structure under study
($T=550^\circ$~C for QD growth and $T=500^\circ$~C for overgrowth by
Si). However, we reduce the amount of deposited Ge down to 4~ML in
each QD layer, and, as result, we obtain a well-ordered array of
nanodisks after overgrowth by Si. Thus, we can compare two
structures: 1) a non-ordered array with groups of closely spaced QDs
and 2) a well-ordered array of nanodisks, one nanodisk instead of
one QD group. The average size of QD groups coincides with the
characteristic size of nanodiscs.

ESR data obtained on the test structure with nanodisks confirm the
model proposed in this work. ESR signal has isotropic g-factor
$g=1.9992\pm0.0001$ and isotropic ESR line width $\Delta
H_{pp}\approx0.4$~Oe.  Absolute value of g-factor is the same as in
the structure with QD groups at $\theta > 30^{\circ}$. This can be
explained by identical electron localization radius and identical
temperature regime of QD creation. The last factor defines the GeSi
intermixing and strain in the QD system, which have a high influence
on g-factor value. The isotropy of ESR line is explained by the
absence of tunneling transitions between nanodisks, which are well
ordered in the plane and positioned at an equal ($\sim100$~nm)
distance from each other. Narrowness of ESR line indicates the high
efficiency of averaging of nuclear magnetic fields by the electron
state with large localization radius and the high uniformity of
array of nanodisks (negligible inhomogeneous broadening). In the
structure with QD groups the averaging by means of tunneling between
dots is not so efficient, then we observe a few times larger ESR
line width.

In summary, we demonstrate  that the existence of closely spaced QD
groups provides the increase of spin relaxation time in QD system.
Changing the electron localization radius by external magnetic field
allows us to catch the effect of ESR line narrowing and obtain at
the special orientation of magnetic field the fourfold increased
time $T_1$ as compared to the recently studied homogeneous QD
arrays.

\begin{acknowledgments}
This work was supported by RFBR (Grants 11-02-00629-a,
13-02-12105,), SB RAS integration project No. 83 and DITCS RAS
project No. 2.5.
\end{acknowledgments}

\end{document}